\documentclass[aps,twocolumn,superscriptaddress]{revtex4-1}
\usepackage{amsfonts}
\usepackage{amssymb}
\usepackage{amsmath}
\usepackage{subfigure}
\usepackage{graphicx}
\usepackage{xcolor}
\usepackage[light]{iwona}
\begin{document}
\title{General linear response formula for non integrable systems obeying the Vlasov equation}
\author{Aurelio Patelli}
\affiliation{CEA - Service de Physique de l'Etat condens\'e, Centre d'Etudes de Saclay, 91191, Gif-sur-Yvette, France} 
\author{Stefano Ruffo}
\affiliation{Dipartimento di Fisica ed Astronomia and CSDC, Universit\`{a} degli Studi di Firenze, CNISM and INFN, Via Sansone 1, 50019 Sesto Fiorentino, Italy} 
%\and the second here
\begin{abstract}
Long-range interacting $N$-particle systems get trapped into long-living out-of-equilibrium stationary states called
quasi-stationary states (QSS). We study here the response to a small external perturbation when such systems are 
settled into a QSS. In the $N \to \infty$ limit the system is described by the Vlasov equation and QSS are mapped into stable 
stationary solutions of such equation. We consider this problem in the context of a model that has recently
attracted considerable attention, the Hamiltonian Mean Field (HMF) model. For such a model, stationary inhomogeneous and
homogeneous states determine an integrable dynamics in the mean-field effective potential and an action-angle 
transformation allows one to derive an exact linear response formula. However, such a result would be of limited interest
if restricted to the integrable case. In this paper, we show how to derive a general linear response formula which does not
use integrability as a requirement. The presence of conservation laws (mass, energy, momentum, etc.) and of further Casimir
invariants can be imposed {\it a-posteriori}. We perform an analysis of the infinite time asymptotics of the response formula
for a specific observable, the magnetization in the HMF model, as a result of the application of an external magnetic field, for two
stationary stable distributions: the Boltzmann-Gibbs equilibrium distribution and the Fermi-Dirac one. When compared
with numerical simulations, the predictions of the theory are very good away from the transition energy from inhomogeneous to
homogeneous states.
\end{abstract}
 %end of abstract
%
\maketitle
\section{Introduction}
\label{Introduction}
The two-body potential of $N$-particle long-range interacting systems decays asymptotically 
as $r^{-\alpha}$, where $r$ is the inter particle distance, $0\le\alpha\le d$ 
and $d$ is the dimension of the embedding space~\cite{review3}.
Systems with long-range interactions show many interesting properties, such as ensemble inequivalence 
and negative specific heat in the microcanonical ensemble~\cite{Leshouches}. 
These properties are a direct consequence of energy non additivity.

The dynamical properties of systems with long-range interactions are also peculiar.
For such systems, the relaxation towards Boltzmann-Gibbs statistical equilibrium 
occurs on a time scale that diverges with system size~\cite{Yamaguchi:2004}. 
This implies that the long-time $t \to \infty$ limit and the thermodynamic $N \to \infty$ limit  
do not commute, because the relaxation time scale $\tau$ diverges with $N$.

On a shorter time scale, $t \ll \tau$, one observes a relaxation towards out-of-equilibrium stationary states. 
There are many such stationary states, that are obtained by varying the initial condition. They have been called 
Quasi Stationary States (QSSs)~\cite{review3,Yamaguchi:2004}, because for all finite $N$ they display a slower 
``collisional" evolution towards equilibrium on a time scale of the order $\tau$. 
On a short time scale, the time evolution of the single-particle distribution function in phase-space
is well described by the Vlasov equation~\cite{Balescu:1997,Nicholson:1992}, with corrections of order $1/N$~\cite{Braun:1977}. 
This equation has an infinity of stationary solutions, the stable ones can give rise to QSSs.

An interesting question, which has been recently investigated~\cite{Patelli,Yamaguchi:2012,Chavanis:2013}, 
concerns the response of a QSS to a stimulus. 
This question can be explored by studying how a stationary stable state of the Vlasov equation reacts to a perturbation. 
This latter can be of different types and origins. 
A first type of perturbation may come from a generic variation of the single-particle distribution function with 
respect to a stationary stable one. 
Typically, this variation vanishes with time due to Landau damping~\cite{Landau:1946}. 
A second type of perturbation consists in applying externally a stochastic noise~\cite{Nardini1,Nardini2}: 
these perturbations generically bring the system towards an equilibrium or non equilibrium steady state,
depending on whether detailed balance is respected or not, on a time-scale which depends on the strength of the noise. 
A third, and final, kind of perturbation consists in adding an external field to the dynamics. 
This latter is the perturbation considered in Refs.~\cite{Patelli,Chavanis:2013} in the context of linear response 
theory for homogeneous states. 
An extension of this approach to non homogeneous states~\cite{Yamaguchi:2012}, based on integrability, 
has led to a characterization of linear response in a more general framework. 
This latter work is based on previous studies of the solutions of the linearized Vlasov equation around 
inhomogeneous states~\cite{Barre:2012}.
Integrability is guaranteed for one-dimensional systems when the unperturbed state is stationary and is present also 
for a few other Hamiltonians in higher dimensions. 
Integrability is implemented by an action-angle transformation, which conveniently separates 
the phase-space variables.

In this paper, we consider perturbations of the dynamics obtained by the addition of a conservative external field.
As in previous work~\cite{Patelli}, we restrict ourselves to the study of small perturbations within linear
response theory. However, we do not use integrability as a necessary requirement to derive the linear response formula. This 
choice has a drawback, since integrals of motion have then to be imposed by hand in order to obtain the correct
result. The Vlasov equation has an infinity of conserved quantities besides those related to
symmetries, the so-called Casimirs~\cite{Morrison:1987}. Since we are unable to impose this infinity of constraints, 
we restrict ourselves to consider a finite number of them (e.g. normalization of the single-particle distribution) and, therefore, derive 
an approximate response formula. The advantage of this approach with respect to previous ones is that our response formula 
can be derived in general also for non integrable systems. Moreover, our method allows us to consider many different
unperturbed states, both homogeneous and inhomogeneous.

In order to illustrate the validity of our approach we explicitly derive the response to an externally applied
field for the Hamiltonian Mean-Field (HMF) model~\cite{Spohn:1982,Inagaki:1993,Ruffo:1995}, a paradigmatic mean-field system.
This allows us to compare our approximate formula with the exact one derived in Ref.~\cite{Yamaguchi:2012}. 
We obtain a good agreement excluding a region of energy close to the second-order phase transition of the model,
where the response diverges.

The paper is organized as follows. In Section~\ref{VlasovEquation}
we introduce the Vlasov equation and its linearization.
In the following Section~\ref{sec4:VariationHamiltonian} we derive the linear response formula.
Its time asymptotics is studied in Section~\ref{sec4:Asymptotics}. 
Section~\ref{sec4:Constraints} is devoted to the discussion of the constraints of the dynamics.
After introducing the HMF model in Section~\ref{HMFmodel}, we derive the linear response formula for this
model in Section~\ref{sec4:Applicaton}.
In the final Section~\ref{sec4:Numerics}, we demonstrate the agreement of our predictions with numerical
simulations realized for different unperturbed states of the HMF model.

%%%%%%%%%%%%%%%%%%%%%%%%%%%%%%%%%%%%%%%%%%%%%%%%%%%%%%%%%%%%%%%%%%%%%%%%%%%%
\section{The Vlasov equation, its linearization and the response formula}
\label{VlasovEquation}
%%%%%%%%%%%%%%%%%%%%%%%%%%%%%%%%%%%%%%%%%%%%%%%%%%%%%%%%%%%%%%%%%%%%%%%%%%%%
For the sake of completeness, we recall in this Section some basic features of the Vlasov equation and of its linearization.
Moreover, we formally derive the expression of the response to a perturbation in a slightly different way as done in Ref.~\cite{Patelli}. 

The dynamics of an isolated $N$-particles system interacting via a two-body smooth potential is well described by the Vlasov equation 
in the limit $N \to \infty$~\cite{Balescu:1997,Nicholson:1992,Braun:1977}. It is therefore convenient to 
begin with a system composed of an infinite number of particles. In one dimension, each particle is described by a pair of conjugate 
variables $(q,p)$. For simplicity, we consider a periodic domain in $q$. These conjugate variables
must be interpreted as Eulerian variables describing a point in phase-space rather than as dynamical variables.
The Vlasov equation describes the time evolution of the single-particle distribution function $f(q,p,t)$,
\begin{equation}
	\frac{\partial}{\partial t} f=-p\frac{\partial}{\partial q}f+\frac{\partial \phi[f]}{\partial q}\frac{\partial}{\partial p}f~, 
\label{eq:Vlasov}
\end{equation}
where
\begin{equation}
	\phi[f](q,t)=\int dxdv\, u(q-x)f(x,v,t)
\label{eq:SelfConsistency}
\end{equation}
is the mean-field potential. In the $N \to \infty$ limit, the two-body correlation function vanishes and, therefore, 
the single-particle distribution function contains all the information about the state of the system.
The Vlasov equation can be rewritten as a Liouville equation by using the Hamiltonian
\begin{equation}
	\mathcal{H}[f](q,p,t)=\frac{p^2}{2}+\phi[f](q,t)~,
	\label{eq:VlasovHamiltonian}
\end{equation}
which is a functional of the single-particle distribution function. 

In the following we will need to define the time evolution of specific observables. To this aim, we define
the Poisson brackets of differentiable single-particle observables $a(q,p)$ and $b(q,p)$ as usual:
\begin{equation}
	\{a,b\}=\frac{\partial a}{\partial q}\frac{\partial b}{\partial p}-\frac{\partial a}{\partial p}\frac{\partial b}{\partial q}~.
\end{equation}
In terms of these brackets the Liouville operator $\mathcal{L}$ acting on $a(q,p)$, can be written as
\begin{equation}
	\mathcal{L}[f](q,p,t)a(q,p)=\{a(q,p),\mathcal{H}[f](q,p,t)\}~.
	\label{eq:Liouvillian}
\end{equation}
In order to implement the constraints of the dynamics in the response formula, it is useful to
remark that the kernel of the Liouville operator~\eqref{eq:Liouvillian} contains at least the 
constant function, that we denote $\delta M$ for applications in the following, the Hamiltonian itself,
$\mathcal{H}$, and all the smooth functions of the Hamiltonian $\mathsf{f}(\mathcal{H})$,
as proved by Jeans~\cite{Jeans:1919}.

Let us consider a stationary solution of the Vlasov equation, $f_0(q,p)$. Following Jeans~\cite{Jeans:1919}, we 
consider the class of solutions that are functions of the Hamiltonian~\eqref{eq:VlasovHamiltonian}
\begin{equation}
	f_0(q,p)=\mathsf{F}(\mathsf{x}(q,p)),\quad \mathsf{x}(q,p)=\sigma \mathcal{H}[f_0](q,p)~,
	\label{eq:JeansDistribution}
\end{equation}
where $\mathsf{F}$ is a smooth function of a single variable and $\sigma$ is
an inverse energy scale. The function $\mathsf{F}$ is such that $f_0$ is indeed 
a distribution, i.e. a non negative and normalizable function.

We now perturb the system by switching on instantaneously an external field $h b(q,t)$ at time $t=0$.
This restriction is adopted for the sake of simplicity, but our approach is applicable, with minor modifications, 
also to the case in which the field is switched on smoothly. The size of the perturbation is vanishingly small $h \ll 1$.

The linearization procedure of the Vlasov equation around the stationary unperturbed solution consists in developing 
in powers of $h$ all the relevant quantities, i.e. the distribution function and the Hamiltonian, and in keeping only the
first-order terms in $h$. In order to verify the validity of this procedure, one can check that in the limit $h \to 0$ 
the unperturbed solution is recovered. We will denote all the unperturbed quantities with the subscript zero, 
e.g. $\mathsf{x}=\mathsf{x}_0+h\delta\mathsf{x}+\mathcal{O}(h^2)$.
The perturbed distribution function can be developed in two different ways 
\begin{eqnarray}
	f(q,p,t)&=&\mathsf{F}(\mathsf{x}_0)+h\frac{d\mathsf{F}}{d\mathsf{x}}(\mathsf{x}_0)\delta \mathsf{x}+\mathcal{O}(h^2)\\
	      &=&f_0(q,p)+h\delta f(q,p,t)+\mathcal{O}(h^2).
	      \label{eq:distributionseries}
\end{eqnarray}
The first development is strictly valid only for instantaneous variations of the external field. For smooth variations
one has to take into account the functional variations of $\mathsf{F}$. We discuss in detail this aspect in 
Appendix~\ref{app4:Energy-Casimir}.
The perturbed Hamiltonian is given by
\begin{equation}
	\mathcal{H}[f]=\mathcal{H}_0[f_0]+hb(q,t)+h\phi[\delta f]+\mathcal{O}(h^2),
	\label{eq:pertHam}
\end{equation}
where $\mathcal{H}_0[f_0]$ is the unperturbed Hamiltonian.
In terms of the two-body potential the variation of the Hamiltonian is given by
\begin{equation}
	\delta \mathcal{H}(q,t)=b(q,t)+\int dxdv \delta f(x,v,t)u(q-x)
	\label{eq:VariationHamiltonian}
\end{equation}
and it is a functional of the variation of the distribution function.
Inserting $f$ and $H[f]$ defined in Eqs.~\eqref{eq:distributionseries} and \eqref{eq:pertHam} into the Vlasov equation \eqref{eq:Vlasov} and
collecting terms of order $h^0$ and $h$ one gets the following set of coupled equations
\begin{eqnarray}
	\frac{\partial}{\partial t}f_0&=&p\frac{\partial}{\partial q}f_0-\frac{\partial \phi[f_0]}{\partial q}
	\frac{\partial}{\partial p}f_0=0~,\\
	\frac{\partial}{\partial t}\delta f&=&p\frac{\partial}{\partial q}\delta f-\frac{\partial \phi[f_0]}
	{\partial q}\frac{\partial}{\partial p}\delta f -\frac{\partial \delta \mathcal{H}}{\partial q}\frac{\partial}
	{\partial p} f_0\label{eq:LinearVlasovEquation}\nonumber\\
	&=&\{\mathcal{H}_0,\delta f\}+\{\delta \mathcal{H},f_0\}~.
\end{eqnarray}
Solving the first of these equations gives the unperturbed distribution function $f_0(q,p)$ which, substituted
into the second equation, which is nothing but the linear Vlasov equation, leads to the Duhamel's formula
for $\delta f$ 
\begin{equation}
	\delta f(q,p,t)=\int_0^t d\tau\, \mathcal{U}(t-\tau) \{\delta \mathcal{H}(\tau),f_0\}~,
	\label{eq:DistrVar}
\end{equation}
where the operator $\mathcal{U}(t)=e^{t\mathcal{L}_0}$ is the evolution operator of the unperturbed dynamics,
with $\mathcal{L}_0$ the Liouvillian defined in Eq.~\eqref{eq:Liouvillian}.
In deriving the last equation, we have set to zero the initial variation of the distribution $\delta f(q,p,0)$ 
since the external perturbation vanishes at $t=0$. However, this does not solve the problem, since we need
to know $\delta  \mathcal{H}$ in order to determine $\delta f$. On the other hand, we do have an equation which gives
$\delta  \mathcal{H}$ in terms of $\delta f$, Eq.~\eqref{eq:VariationHamiltonian}. Therefore, inserting $\delta f$ given
in Eq.~\eqref{eq:DistrVar} into Eq.~\eqref{eq:VariationHamiltonian}, we obtain a closed equation for $\delta \mathcal{H}$
\begin{eqnarray}
	\delta \mathcal{H}(q,t)&=&b(q,t)+\label{eq:HamEq}\\
	&\,&+\int_0^td\tau \int dxdv\,u(q-x)\mathcal{U}(t-\tau)\{\delta \mathcal{H}(x,\tau),f_0\}\nonumber
\end{eqnarray}
In Section~\ref{sec4:VariationHamiltonian} we will obtain a formal solution of this equation using Laplace-Fourier techniques. This will
give us in turn the formal solution of the linear Vlasov equation.

The linear Vlasov equation can be solved for stationary homogeneous distributions using Laplace transform 
in time and Fourier transform in space $q$~\cite{Nicholson:1992,Balescu:1997}. Applying these well-known methods, some authors
\cite{Patelli,Yamaguchi:2012} derived explicit analytical formulas for the variation of a generic observable $a(q,p)$
under the application of an external perturbation. These derivations are based on the analogy between the Vlasov
equation and the Liouville equation. Specifically, a Kubo-like \cite{KuboJPSJ:1957} formula was obtained
\begin{eqnarray}
	\delta\langle a\rangle&=&\langle a\rangle_{t}-\langle a\rangle_{0}=h\int_0^t\mathcal{R}(t,s)ds+\mathcal{O}(h^2)\\
	\mathcal{R}(t,s)&=&\langle\{a(t-s),\delta \mathcal{H}(s)\} \rangle_0
	\label{eq:Response}
\end{eqnarray}
where the average  $\langle \cdot \rangle_t$ in the first equation is taken with respect to the distribution at time $t$.
The response function $\mathcal{R}(t,s)$, appearing at the linear order, is defined in the second equation
in terms of the Poisson bracket of the observable $a$ with the variation of the Hamiltonian, averaged with
respect to the initial distribution, $\langle \cdot \rangle_0$. The observable $a(t)=\exp (t\mathcal{L}_0)a(q,p)$ is evolved in time 
using the unperturbed dynamics.

At variance with Kubo's linear response theory at thermodynamical equilibrium, the response function $\mathcal{R}(t,s)$
explicitly depends on two times, i.e. it is not invariant under time translations. We will discuss in the following 
cases in which the system relaxes asymptotically to a stationary state and, therefore, time translation invariance
is restored in this limit. However, there are initial conditions for which the system sets into a stable macroscopically
oscillating state, thus violating time translation invariance for all times. We believe that the presence of
such oscillatory states is a typical behavior of mean-field systems: it is indeed due to the self-consistent interaction
of the single particle with the mean-field potential~\cite{Pakter:2013}.

We have up to now presented in some detail the basic equations which will allow us to derive, at linear order, the
response of the system to a small external perturbation. We will discuss in the following Sections how the calculation 
of the variation of a generic observable can be performed explicitly for a wide class of models and of initial
conditions. However, before moving to this derivation, we would like first to discuss the role of the constraints
on the dynamics imposed by global conservation laws.

%-------------------------------------------------------------------------
\section{Derivation of the response formula}
\label{sec4:VariationHamiltonian}
In the previous Section \ref{VlasovEquation} we have obtained the linear Vlasov equation \eqref{eq:LinearVlasovEquation}
for a general initial distribution and, using Duhamel formula, we have derived closed equations for both the
variation of the distribution function $\delta f$ and the variation of the Hamiltonian $\delta \mathcal{H}$.
In the method we will develop below, we exploit the fact that $\delta f$ does not depend on momentum.

The equation which determines $\delta f$, Eq.~\eqref{eq:HamEq}, is an inhomogeneous integro-differential equation.
Obtaining a solution of this equation allows one to derive $\delta f$ using Duhamel's formula~\eqref{eq:DistrVar}.

Laplace transforming both sides of \eqref{eq:HamEq} yields
\begin{eqnarray}
	\delta\hat{ \mathcal{H}}(q,\omega)&=&\hat b(q,\omega)+ \nonumber \\
        &&+\int dxdv\, u(q-x)\mathrm{R}(\omega)\{\delta \hat{\mathcal{H}}(x,\omega),f_0\},\quad\quad\label{eq:VariationalHamiltonianLaplace}\\
	\mathrm{R} (\omega)&=&\mathrm{L}[\mathcal{U}](\mathrm{\omega})=\int_0^\infty e^{-\imath\omega t}e^{t\mathcal{L}_0}dt=\frac{1}{\imath\omega-\mathcal{L}_0},\label{eq:Resolvent}
\end{eqnarray}
where the hat identifies Laplace transformed functions and the operator $\mathrm{R}$ is the Laplace 
transform of the evolution operator $e^{t\mathcal{L}_0}$.
We consider Jeans' distributions \eqref{eq:JeansDistribution} as initial states, which appear as the most natural ones.

The Hille-Yosida theorem~\cite{ReedSimon} states that the Laplace transform of an operator, 
defining a semi-group by the generator $\mathcal{L}_0$, is the resolvent of the generator itself. 
A general property of resolvents, such as \eqref{eq:Resolvent}, is that they satisfy the identity
\begin{equation}
	\mathrm{R}\mathcal{L}_0=\imath\omega\mathrm{R}-\mathrm{I}-[\mathrm{R},\mathcal{L}_0]~.
	\label{eq:ResProperty}
\end{equation}
In the following, we will assume that the commutator between the Liouville operator and its resolvent is zero 
in order to avoid technical problems.
The dependence of the initial distribution on the unperturbed Hamiltonian transforms the Poisson bracket in 
Eq.~\eqref{eq:VariationalHamiltonianLaplace} into the Liouville operator
\begin{equation}
	\{\delta\mathcal{H},f_0\}=\sigma_0 f_0'\mathcal{L}_0\delta\mathcal{H},
\end{equation} 
where
\begin{equation}
	f_0'(q,p)=\frac{d \mathsf{F}}{d\mathsf{x}}\Big|_{\mathsf{x}=\sigma_0\mathcal{H}_0(q,p)}~.
\end{equation}
With $\mathrm{x}$ we identify the argument of the Jeans' distribution, which corresponds to an adimensional
quantity, $\mathrm{x}=\sigma_0\mathcal{H}_0$, where $\sigma_0$ is an inverse energy scale that
can depend on some macroscopic parameter.
For instance, at equilibrium the inverse energy is equal to the inverse temperature, times some constant.
The equation for the variation of the Hamiltonian in the complex Laplace space can be written in the following compact form
\begin{equation}
	\delta\hat{\mathcal{H}}(q,\omega)=\hat b(q,\omega)+\chi[\delta\hat{ H}](q,\omega),
	\label{eq:EqHamLaplace}
\end{equation}
where the operator $\chi$ is defined as
\begin{eqnarray}
	\chi[\varphi](q,\omega)&=&\int dx dv \, \sigma_0 u(q-x)f_0'\mathrm{K}(\omega) \varphi(x),\\
	\mathrm{K}(\omega)&=&-1+\imath\omega\mathrm{R}(\omega),
\end{eqnarray}
and $\varphi(x)$ is a generic test function.

A feature of homogeneous initial states is that Fourier modes are decoupled 
and equation \eqref{eq:EqHamLaplace} can be solved using the Fourier transform.
In this framework, the modes of the variation of the Hamiltonian are
\begin{equation}
	\tilde{\delta\mathcal{H}}(k,\omega)=\frac{\tilde{b}(k,\omega)}{\epsilon(k,\omega)}~,
\end{equation}
where $\epsilon(k,\omega)$ is the dielectric function \cite{Nicholson:1992,Landau:1946}.

On the contrary, inhomogeneous initial states couple all the modes 
and Eq.~\eqref{eq:EqHamLaplace} cannot be simplified using the Fourier transform.
Therefore, we are led to assume that a base of eigenfunctions $\varphi_i(q), i \in \mathrm{N}$ exists, 
which solves the eigenvalue equation
\begin{equation}
	\chi[\varphi_i](q,\omega)=\lambda_i(\omega)\varphi_i(q)~,\qquad i=1,\cdots,N
	\label{eq:EigenvalueEquation}
\end{equation}
where $\lambda_i$ is the associated eigenvalue.
This base plays the role of the Fourier base in the inhomogeneous regime.
Therefore, we consider external potentials that belong to the subspace of functions spanned by these eigenfunctions
\begin{equation}
	\hat{b}(q,\omega)=\sum_i a_i(\omega)\varphi_i(q), \quad q_i(\omega)\in\mathbb{C}
	\label{eq:PotentialSeries}
\end{equation}
A basic tool to solve integral equations is to use the Liouville-Neumann series~\cite{KatoPertSeries} which is a recursive formula.
This allows us to obtain $\delta\hat{\mathcal{H}}$ and, in turn, the response \eqref{eq:Response}. In our case,
the Liouville-Neumann series reads
\begin{equation}
	\delta\tilde{\mathcal{H}}(q,\omega)=\hat{b}(q,\omega)+\sum_{n=1}^\infty \chi^n[\hat b](q,\omega)~,
	\label{eq:LiouvilleNeumann}
\end{equation}
where $\chi^n$ is the $n$th iterate of $\chi$.
Substituting the definition of the external potential in terms of the eigenfunctions into 
Eq.~\eqref{eq:LiouvilleNeumann} and resumming the geometric series one gets
\begin{equation}
	\delta \hat{\mathcal{H}} = \sum_{i \in \mathrm{N}} a_i(\omega)\frac{\varphi_i(q)}{1-\lambda_i(\omega)},\quad |\lambda_i|<1~.
	\label{eq:LNsolution}
\end{equation}
The variation of the Hamiltonian which solves Eq.~\eqref{eq:EqHamLaplace} can be expressed in terms of the same 
eigenfunctions of the external potential but with a different amplitude.
That solution has some properties related to the ones of the homogeneous regime.
For instance, by analogy, we can denote the denominator of \eqref{eq:LNsolution} as
\begin{equation}
	\varepsilon_i(\omega)=1-\lambda_i(\omega),
	\label{eq:DielectricFunctionReal}
\end{equation}
which is nothing but a dielectric-like function for inhomogeneous states.
Whenever this function is zero in the complex-$\omega$ space, $\delta\hat{\mathcal{H}}$ has a pole.
Therefore, zeros of $\varepsilon_i(\omega)$ characterize the long-time behavior of $\delta \hat{\mathcal{H}}$,
after performing an inverse Laplace transform, as we will discuss in detail in the following Section.
Moreover, when the system is homogeneous, formula~\eqref{eq:DielectricFunctionReal} 
reduces to the formula for the dielectric function and the eigenfunctions $\varphi_i$ 
become the Fourier modes.
Furthermore, we obtain a stability criterion because the geometric series can be computed only when $\vert\lambda_i\vert<1$.
On the contrary, when the eigenvalue is larger than one in modulus the state is unstable and linear theory cannot be used.
We remark that integral equations can also be solved using Fredholm determinants~\cite{Whittaker}, but in this case
the solution is more involved, since the kernel of the integral is itself a non-trivial operator.

%-----------------------------------------------------------------------------------------------------------------
\section{Time asymptotic behaviour}
\label{sec4:Asymptotics}
Laplace {\it Limit Theorem} relates the asymptotic value of a function $g(t)$ at $t\to\infty$ to the limit
of its Laplace transform $\hat{g}(\omega)$ for $\omega\to0$.

Let us denote every quantity evaluated in this regime with the index $\infty$, e.g., the external potential
\begin{equation}
	\hat b_{\infty}(q)=\lim_{\omega\to0^-}\imath\omega\hat b(q,\omega)= \lim_{t\to\infty}\hat b(q,t).
\end{equation}
The Liouville-Neumann series for asymptotic times reads
\begin{equation}
	\delta \hat{\mathcal{H}}_{\infty}(q)=b_\infty(q)+\sum_{i=1}^\infty\chi_\infty[b_\infty](q),
	\label{eq:AsyLN}
\end{equation}
and the limit of the eigenvalue equation \eqref{eq:EigenvalueEquation} is
\begin{eqnarray}
	\lambda_i\varphi(q)&=&-\sigma_0\int dxdv f'_0(x,v)u(q-x)\varphi(x) + \nonumber\\
	&\,&+\imath\sigma_0\int dxdv f'_0u(q-x)\lim_{\omega\to0}\omega\mathcal{R}(\omega)\varphi(x).
	\label{eq:EigenEquation}
\end{eqnarray}
The first integral does not depend on the Laplace variable $\omega$ while the second 
one depends on $\omega$ through the kernel operator $\omega\mathcal{R}(\omega)$.
This latter integral converges to zero when the initial state is homogeneous, because its Fourier transform is
$\omega(\epsilon(k,\omega)-1)$ and in the limit $\omega\to0$ it results in a vanishing contribution.
On the contrary, when the initial state is inhomogeneous the integral could converge to a finite value in the same limit.

Let us consider a function $g(x,v)$: when it is in the kernel of the Liouville operator \eqref{eq:Liouvillian}
the action of the resolvent becomes trivial and the limit $\omega\to0$ gives the identity
\begin{eqnarray}
	\lim_{\omega\to0} \omega\mathcal{R}(\omega)g(x,v)=g(x,v).
\end{eqnarray}
Consequently, in the time asymptotic, the operator $\mathrm{K}(0)$  gives zero when evaluated on these functions.

The operator $\mathcal{L}_0$ has an empty continuum spectrum also on some manifolds
in the phase space and these manifolds could give a finite contribution to the response.
The Fourier transform of the second integral in Eq.~\eqref{eq:EigenEquation} is
\begin{equation}
	\mathcal{J}_\mathsf{\omega}=\int dxdvf_0'(\mathcal{H}_0)e^{\imath kx}\frac{\omega}{\omega-\imath\mathcal{L}_0}\varphi(x),
	\label{eq:ManifoldIntegral}
\end{equation}
and in the asymptotic time limit it gives a non zero integrand whenever the larger (in modulo) eigenvalue of $\mathcal{L}_0$
scales as $\omega$.
We call $\mathcal{M}$ the manifold in which the Liouville operator is identically zero, thus the manifold where the
integrand could get a contribution in the limit.
Unfortunately, we don't know how to obtain this manifold in general and even if it really exists.
We will discuss in more detail formula \eqref{eq:ManifoldIntegral} in Appendix \ref{app4:TheSecondIntegral} and
hereafter we assume that its contribution can be negligible, since in the homogeneous phase it is indeed zero.

Discarding the second integral in Eq.~\eqref{eq:EigenEquation}, the operator $\mathrm{K}$ turns out to be proportional to the identity
and the other quantities become
\begin{subequations}
	\begin{eqnarray}
		\mathrm{K}(0)&=&-1\\
		\lambda_i\varphi_i(q)&=&\sigma_0\int u(q-x) f_0'\varphi_i(x)\\
		\delta  \hat f_{\infty}(q,p)&=&\sigma_0\delta \hat{ \mathcal{H}}_{\infty}(q)f_0'(q,p)
	\end{eqnarray}
\end{subequations}
Let us make two short remarks. The first remark is that the response~\eqref{eq:Response} at every time becomes ill defined.
Let us then introduce the integrated response for asymptotic times, defined as
\begin{subequations}
	\begin{eqnarray}
		\delta \bar a&=&\lim_{t\to\infty}\frac{\partial}{\partial h}\Big(\langle a\rangle_t-\langle a \rangle_0\Big)\Big|_{h=0},\\
		&=&\int dpdq\,\delta f_\infty(q,p) a(q,p).
	\end{eqnarray}
	\label{eq:AsyIntResp}
\end{subequations}
This formula describes the finite variation of a single-particle observable $a(x,v)$ at the linear order in
the perturbation parameter $h$.

The second remark concerns the stability criterion of the initial state.
From the geometric series~\eqref{eq:LNsolution}, we get a criterion to perform the resummation: $|\lambda_i|<1$ for all $i$.
In the asymptotic time regime this inequality can be seen as a criterion for which the initial state is stable,
since otherwise the series diverges.
It can be compared with other stability criteria \cite{Penrose:1959,StabilityShun}.

%-------------------------------------------------------------------------
\section{Constraints}
\label{sec4:Constraints}
Vlasov dynamics is characterized by the existence of constraints of different nature~\cite{Holm:1984}. 
We will here restrict to discuss those constraints that play a relevant role in the derivation of our response theory. 
The first constraint derives from mass conservation, which is a consequence of considering a closed system. 
However, our system can exchange energy with the environment, hence we can evaluate the work done on 
the system by the external perturbation. 
The entailed generalized energy conservation relation imposes, as we will see, a constraint on the dynamics. 
Isolated systems also conserve total momentum. 
However, one can choose perturbations of the dynamics which determine variations $\delta f(q,p,t)$ 
that are even functions of $p$.
This in turn implies that momentum is necessarily conserved and can for convenience be set to zero. 
The effect of the conservation of other global invariants associated to symmetries, like angular momentum, 
will not be discussed in this paper, since the systems we consider are one-dimensional.

On top of that, the Vlasov equation is endowed with an infinity of conserved quantities, the so-called 
Casimirs~\cite{Morrison:1987,Holm:1984}. 
They all take the form $\int c(f(x,v)) dx dv$, with $c$ a smooth function. 
For instance, mass is one of such Casimirs. 
We will assume that the effects induced by all these conservations, besides mass, are negligible.

\subsection{Mass constraint}
The mass of the system is given by
\begin{equation}
	M(t)=\int f(x,v,t) \, dx dv~.
\end{equation}
At first order in the size of the perturbation $h$ we get
\begin{equation}
	M(t)=M(0)+h\int\delta f\,dxdv+\mathcal{O}(h^2).
\end{equation}
For a closed system, mass is conserved $M(t)=M(0)$, hence
\begin{equation}
	\int \delta f(x,v,t) \,dxdv=0~.
	\label{eq:MassConstraintAT}
\end{equation}
In the following we will set the total mass to one, $M=1$.

\subsection{Energy constraint}
The system is perturbed by an external field, hence the total energy 
\begin{eqnarray}
	E(t)&=&\int \mathrm{E}[f](x,v,t)f(x,v,t)dxdv\\
	\mathrm{E}[f](q,p,t)&=&\frac{p^2}{2}+\frac{1}{2}\phi[f](q,t)+hb(q,t)~.
\end{eqnarray}
is not a conserved quantity. 
However, since the time evolution of the distribution function is determined by the full perturbed Hamiltonian 
(the external forces are conservative), the variation of the total energy is equal to the work $W(t)$ done by 
the external forces on the system
\begin{equation}
	E(t)-E_{0}=W(t)~,
	\label{eq:EnergyConstraintDef}
\end{equation}
where
\begin{equation}
	W(t)=h\int f(x,v,t)b(x,t)dxdv~.
\end{equation}
This is a generalized energy conservation law, as obtained in mechanics.

Let us remark that the functional $\mathrm{E}[f]$ is the phase-space observable associated
with the total energy and differs from the Hamiltonian $\mathcal{H}[f]$ by the factor $1/2$ in front of
the mean-field potential. 
This avoids the double counting of the interaction energy between two separated regions in phase-space.

At linear order in $h$ one gets the following constraint on the variation of the distribution function
\begin{equation}
	\int \mathcal{H}_0[f_0](x,v)\delta f(x,v,t)dxdv=0,\quad \forall t>0,
	\label{eq:EnergyConstraintAT}
\end{equation}
where we have used the identity 
\begin{equation}
\int f\phi[\delta f]dqdp=\int \delta f\phi[f]dqdp.
\end{equation}

\subsection{Implementation of the constraints}
The asymptotic stationary state is described by the stationary linear Vlasov equation
\begin{equation}
	\left(p\frac{\partial}{\partial q}-\frac{\partial \phi_0}{\partial q}\frac{\partial}{\partial p}\right)\delta f_\infty + \frac{\partial \delta \mathcal{H}}{\partial q}\frac{\partial}{\partial p}f_0=0.
	\label{eq:StationaryVlasovLinear}
\end{equation}
The formal solution of this equation is
\begin{equation}
	\delta f(q,p)=f_0'(q,p)\delta\mathsf{x}(q,p),
	\label{eq:AsyDeltaF}
\end{equation}
and it corresponds to the Duhamel formula \eqref{eq:DistrVar} for large times.
The variation $\delta\mathsf{x}$ cannot a priori be defined because the kernel of the 
Liouville operator \eqref{eq:Liouvillian} is not null.
For example, the function
\begin{equation}
	\delta\mathsf{x}=\sigma_0\delta \mathcal{H}+\mathcal{H}_0\delta\sigma+\delta M,
	\label{eq:DeltaX}
\end{equation}
with $\delta M, \delta\sigma \in\mathbb{R}$ two arbitrary constants, gives a solution of the linear stationary Vlasov equation.
These new terms are useful to implement the constraints of the system.
For instance, the two equations \eqref{eq:MassConstraintAT} and \eqref{eq:EnergyConstraintAT} become
\begin{eqnarray}
	\int f_0'(\sigma_0 \delta \mathcal{H}+\mathcal{H}_0\delta\sigma+\delta M)\mathcal{H}_0\,dqdp=0,\label{eq:EnergyConstraint}\\
	\int f_0'(\sigma_0 \delta \mathcal{H}+\mathcal{H}_0\delta\sigma+\delta M)\,dqdp=0,\label{eq:MassConstraint}
\end{eqnarray}
where the phase space dependencies are dropped for the sake of clarity.
In order to solve this linear problem we define
\begin{subequations}
	\label{eq:ConstraintsFunctions}
	\begin{eqnarray}
		J_{\mathcal{H}_0^2}&=&\int f_0'  \mathcal{H}_0^2\,dqdp,\\
		J_{\mathcal{H}_0}&=&\int f_0' \mathcal{H}_0\,dqdp,\\
		J_{1}&=&\int f_0'\,dqdp,\\
		J&=&J_{\mathcal{H}_0^2}J_1-J_{\mathcal{H}_0}^2,
	\end{eqnarray}
\end{subequations}
then equations \eqref{eq:EnergyConstraint} and \eqref{eq:MassConstraint} give the following functional relations
\begin{eqnarray}
	\delta M[\delta  \mathcal{H}]=\sigma_0\frac{J_{\mathcal{H}_0}\int f_0' \mathcal{H}_0\delta\mathcal{H}-J_{\mathcal{H}_0^2}\int f_0'\delta \mathcal{H} }{J},\\
	\delta\sigma[\delta \mathcal{H}]=\sigma_0\frac{J_{\mathcal{H}_0}\int f_0'\delta\mathcal{H}-J_{1}\int f_0' \mathcal{H}_0 \delta \mathcal{H}}{J}.
\end{eqnarray}
The parameters $\delta M$ and $\delta\sigma$ take into account the variation of the mass of the system and the variation of
the inverse energy scale, due to the perturbation.
They are linear functionals with respect to their arguments, here the variation of the Hamiltonian, 
and add more terms to equation \eqref{eq:VariationalHamiltonianLaplace}.
Therefore, at infinite times the equation for the variation of the Hamiltonian with two constraints becomes
\begin{eqnarray}
	\delta \mathcal{H}_\infty(q)&=&b+\sigma_0\int u(q-x)f_0'\delta{\mathcal{H}}_\infty+\nonumber\\
		&\,&\delta\sigma[\delta \mathcal{H}_\infty]\int f_0' u(q-x)\mathcal{H}_0+\nonumber\\
		&\,&\delta M[\delta\mathcal{H}_\infty]\int f_0'u(q-x).
\end{eqnarray}
The linearity property of these new terms preserves the Liouville-Neumann form of the solution, but
with a modified eigenvalue equation, which reads
\begin{eqnarray}
	\lambda_i \varphi_i(q)&=&\sigma_0\int u(q-x)f_0'\varphi_i+\nonumber\\
	&\,&\delta\sigma[\varphi_i]\int u(q-x)f_0'\mathcal{H}_0+\nonumber\\
	&\,&\delta M[\varphi_i]\int u(q-x)f_0'
\end{eqnarray}
Although the form of the solution \eqref{eq:LNsolution} is the same, the eigenvalues
change their value by imposing the constraints \eqref{eq:EnergyConstraint} and \eqref{eq:MassConstraint}.
We show in the next Section \ref{sec4:Applicaton} this property for the HMF model.

The case in which the system has only the mass constraint can be computed by imposing $\delta \sigma=0$.
The linear functional $\delta M$ related to the variation of the mass is
\begin{equation}
	\delta M[\delta \mathcal{H}]=-\sigma_0\frac{\int f'\delta \mathcal{H}}{J_1}
	\label{eq:NormDeltaF}
\end{equation}
and the eigenvalue equation now reads
\begin{equation}
	\lambda_i\varphi_i=\sigma_0\int u(q-x)f_0'\varphi_i+\delta M[\varphi_i]\int u(q-x)f_0'
	\label{eq:KernellMassOnly}
\end{equation}

On top of that, for homogeneous initial states, the mean field $\phi_0$ is zero and both the unperturbed Hamiltonian $\mathcal{H}_0$
and the initial distribution function $f_0$ do not depend on space.
In equations \eqref{eq:EnergyConstraint} and \eqref{eq:MassConstraint} only the variation
of the Hamiltonian $\delta \mathcal{H}$ depends on the space variable $q$ and usually its integral over space is zero.
Thus, the two equations of the constraints can be satisfied only with $\delta\sigma=0$
and $\delta M=0$, independently of the initial distribution $f_0$.
Therefore, both the constraints are automatically fulfilled for homogeneous systems.

%%%%%%%%%%%%%%%%%%%%%%%%%%%%%%%%%%%%%%%%%%%%%%%%%%%%%%%%%%%%%%%%%%%%%%%%%%%%
\section{The HMF model}
\label{HMFmodel}
%%%%%%%%%%%%%%%%%%%%%%%%%%%%%%%%%%%%%%%%%%%%%%%%%%%%%%%%%%%%%%%%%%%%%%%%%%%%
In this section we consider a toy model of long-range interacting systems,
the Hamiltonian Mean Field (HMF) model~\cite{Spohn:1982,Inagaki:1993,Ruffo:1995}.
The model is defined by its finite $N$ dynamics but it has a Vlasov counterpart. 
Despite its simplicity, the HMF model captures many of the features of long-range interactions,
including QSS. We begin by discussing its statistical equilibrium properties, in
order to then compare the the response in thermal equilibrium with the Vlasov one.

The HMF model describes $N$ particles moving on the unit circle which interact via a cosine potential
\begin{equation}
	u(q)=1-\cos(q), \quad q\in[0,2\pi)
\label{cosine}
\end{equation}
The canonical coordinate $q$ is an angle and specifies the position of the particle, while $p$ is its conjugate momentum.
The finite $N$ Hamiltonian is
\begin{eqnarray}
	H=\frac{1}{2}\sum_{i=1,N} p_i^2+\frac{1}{2N}\sum_{i,j<N}[1-\cos(q_i-q_j)]
	\label{eq:HMFNham}
\end{eqnarray}
where the $1/N$ factor in front of the second term is such that the potential energy and the kinetic 
energy scale equally with $N$. It is sometimes called in literature the Kac prescription~\cite{Kac:1959}.
The Hamiltonian~\eqref{eq:HMFNham} converges to the Vlasov Hamiltonian~\eqref{eq:VlasovHamiltonian} 
in the mean-field limit $N \to \infty$ with the mean-field potential
\begin{eqnarray}
	\phi[f](q,t)&=&1-m_x\cos(q)-m_y\sin(q)\\
	m_x&=&\int dxdv\, \cos(x)f(x,v,t)\\ 
	m_y&=&\int dxdv\, \sin(x)f(x,v,t).
\end{eqnarray}
where $m_{x},m_{y}$ are the magnetizations along the $x$ and $y$ axes and depend on the distribution function 
which solves the Vlasov equation.
The energy in the mean-field limit is
\begin{eqnarray}
	E&=&\int dqdp\,\mathrm{E}(q,p,t)f(q,p,t)\\
	&=&\int dqdp\,\frac{p^2}{2}f\,+\frac{1}{2}(1-m^2).
\end{eqnarray}
where $m=\sqrt{m_x^2+m_y^2}$, the modulus of the magnetization is the order parameter.
The HMF model shows a second order phase transition in the microcanonical ensemble~\cite{review3} 
from a low energy inhomogeneous phase to a high energy homogeneous phase at the transition energy
$E_t=3/4$. The order parameter $m$ vanishes in the homogeneous phase and is non zero in the inhomogeneous
detecting the tendency of the particles to form a cluster with a given mass profile. In the canonical
ensemble one gets an equivalent prediction, a second order phase transition at the inverse temperature
$\beta_t=2$.

The Boltzmann-Gibbs (BG) equilibrium phase space distribution is
\begin{equation}
	f_{eq}(q,p)=\frac{1}{Z(\beta)}\exp\Big\{-\beta H[f_{eq}]\Big\}
	\label{eq:EquilibriumDistribution}
\end{equation}
where $Z(\beta)$ is the partition sum and $\beta$ is the inverse temperature.
The one-to-one relation between energy in the microcanonical ensemble and inverse temperature
in the canonical ensemble is obtained by solving the following equations
\begin{eqnarray}
	E(\beta)&=&\frac{1}{2\beta}+\frac{1}{2}(1-m^2(\beta))\\
	m(\beta)&=&\frac{I_1(\beta m(\beta))}{I_0(\beta m(\beta))}\label{eq:MagnEq}
\end{eqnarray}
where $I_{0,1}$ are the modified Bessel's functions of order $0,1$, respectively. A straightforward
interpretation of the previously introduced parameter $\sigma_0$ consists in its identification
with $\beta$: $\beta=\sigma_0(E)$, obtained by solving Eqs.~\eqref{eq:MagnEq}.

The BG distribution is of paramount importance because systems with a finite number of particles 
converge towards this distribution as time evolves, irrespective of the initial state.
BG equilibrium is also a stationary stable solution of the Vlasov equation~\eqref{eq:Vlasov}, 
thus it can also be considered as a QSS. BG equilibrium is also well defined for inhomogeneous 
states with external fields. In this latter case, magnetization solves the following equation
\begin{equation}
	m'(\beta,h)=\frac{I_1}{I_0}\Big(\beta(m'(\beta,h)+h)\Big)~.
\end{equation}
If we develop this consistency equation in a Taylor series around the unperturbed state $h=0$, we get 
\begin{equation}
	\delta m=\frac{1/\beta+I_2/I_0-m^2}{m^2-I_2/I_0}.
	\label{eq:ThermoCanonicalResp}
\end{equation}
We call this reaction to the application of the field ``canonical response".

Let us now consider the microcanonical ensemble. Due to the different constraints on the system the 
thermodynamic response cannot be the same as for the canonical ensemble.
The inverse energy scale, given by microcanonical inverse temperature $\beta$, changes under the
action of the perturbation as
\begin{equation}
	\beta\to \beta'=\beta+h\delta \beta+\mathcal{O}(h^2).
\end{equation}
Using the energy constraint we relate $\delta\beta$ to the unperturbed magnetization $m_0$ by
\begin{equation}
	\delta \beta=-2m_0\beta^2\delta m~,
\end{equation}
and the thermodynamic response of the magnetization becomes
\begin{equation}
	\delta m=\beta\frac{1/\beta+I_2/I_0-m^2}{1-\beta(1/\beta+I_2/I_0-m^2)(1-2m^2\beta)}~.
	\label{eq:MicroCanThermo}
\end{equation}
We denote it by ``microcanonical response".

%-------------------------------------------------------------------------
\section{Response formula for the HMF model}
\label{sec4:Applicaton}
Let us consider the eigenvalue equation~\eqref{eq:EigenvalueEquation} of a system that is initially at 
equilibrium or in a Fermi-Dirac QSS~\cite{Patelli}.
The long-range potential of the HMF model is a cosine, see Eq.~\eqref{cosine}. Then, by using the addition formula 
of trigonometries functions, we derive that the eigenfunctions $\varphi_i$ must be sums of sines and cosines.
We consider the base composed by functions that are ``parallel" or ``perpendicular" with respect to the 
spontaneous unperturbed magnetization $\vec{m}=(m_x,m_y)$
\begin{eqnarray}
	\varphi_1(q)&=&\alpha_1\left(\cos(q)+\frac{m_y}{m_x}\sin(q)\right),\\
	\varphi_2(q)&=&\alpha_2\left(\frac{m_y}{m_x}\cos(q)+\sin(q)\right),
\end{eqnarray}
where 	$\alpha_1,\alpha_2\in\mathbb{R}$ are two constants defined by the orthogonality relations
\begin{equation}
	\int dxdv\, \varphi_i(x)\varphi_j(x)f_0'(x,v)=\delta_{i,j},\quad i,j=1,2.
\end{equation}

Invariance under angle translations of the HMF model in the unperturbed state allows us to define 
arbitrary $x$ and $y$ directions. We are therefore free to identify with $x$ the direction of the spontaneous 
magnetization; the magnetization components then become $m_y=0$ and $m_x=|\vec{m}|=m$ and the two eigenfunctions are
\begin{equation}
	\varphi_1(q)=\alpha_1\cos(q), \quad \varphi_2=\alpha_2\sin(q).
\end{equation}

\begin{figure}[t!]
	\centering
	\includegraphics[width=85mm]{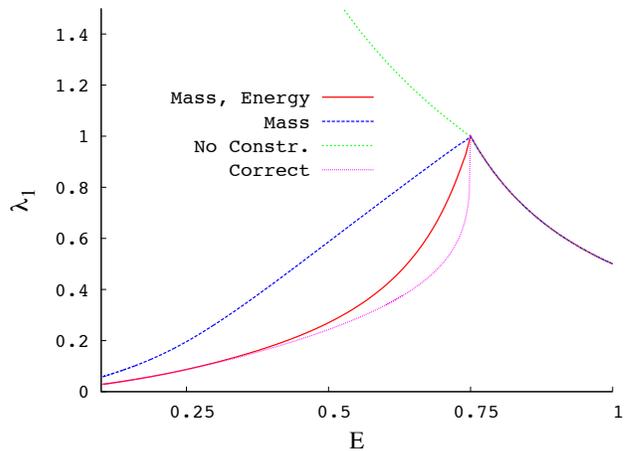}
	\caption{\small{Eigenvalue $\lambda_1$ versus energy $E$ for the HMF model~\eqref{eq:HMFNham} perturbed
            around the equilibrium state~\eqref{eq:EquilibriumDistribution} for the unconstrained case~\eqref{eq:Equilibrium}
		(dotted green line), with only the mass constraint~\eqref{eq:LMass} (dotted blue line) and with both
		mass and energy constraint~\eqref{eq:LEnergyMassConstr} (full red line). The full purple line reproduces
            the exact result obtained in Ref.~\cite{Yamaguchi:2012}. All theories give $\lambda_1=1$ at the phase
		transition energy $E_t=3/4$.
	}}
\label{fig:EqLambda}
\end{figure}

Without loss of generality, we put the external field in the direction of the spontaneous magnetization $x$,
because we are not interested in studying the Goldstone modes associated with rotations of the magnetization
vector. An external field applied perpendicularly to the spontaneous magnetization direction 
excites modes that persist for indefinite time, due to the rotational symmetry of the unperturbed Hamiltonian.
With $\alpha_{1}=1$, the eigenvalue $\lambda_1$ can be obtained from the following formulas
\begin{eqnarray}
	\lambda_1&=&\beta\Phi[f_0'\varphi_1],\label{eq:EigenvalueNoConstraints}\\
	\Phi[r]&=&-\int dxdv\, \varphi(x) r(x,v).
\end{eqnarray}

In the next Subsections we will consider the solution of the eigenvalue equation for the equilibrium state in three cases: 
absence of constraints, imposing the mass constraint and requiring both the mass and the energy constraints. 
Later on, we will discuss the eigenvalue equation for the Fermi-Dirac distribution for the three cases.
\begin{figure}[t!]
	\centering
	\includegraphics[width=85mm]{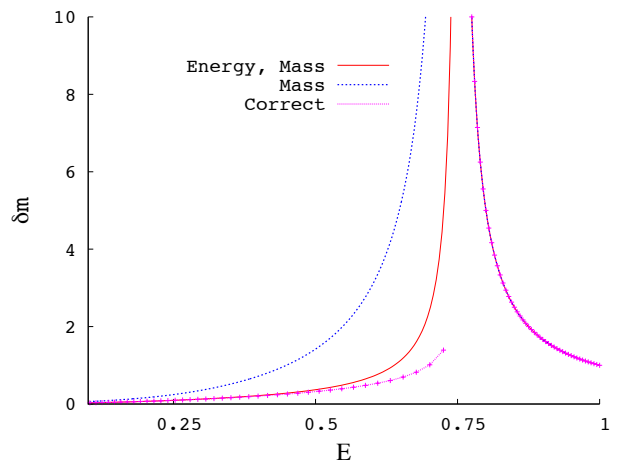}
	\caption{\small{Response $\delta m$ vs. energy $E$ for the HMF model~\eqref{eq:HMFNham} perturbed
            around the equilibrium state~\eqref{eq:EquilibriumDistribution} for the case with mass constraint
		\eqref{eq:ThermoCanonicalResp} (dotted blue line) and for the case with mass and energy constraint
		\eqref{eq:MicroCanThermo} (full red line). The full purple line is
            the exact result obtained in Ref.~\cite{Yamaguchi:2012}. The response diverges at the critical 
		energy $E_t=3/4$.
	}}
	\label{fig:Resp}
\end{figure}

\subsection{Equilibrium without constraints}
The eigenvalue of the sine function is zero, $\lambda_2=0$, because the distribution is even in $q$.
On the contrary, along the $x$ direction we get
\begin{equation}
	\lambda_1=\beta\frac{I_1(\beta m)+\beta mI_2(\beta m)}{\beta mI_0(\beta m)},
	\label{eq:Equilibrium}
\end{equation}
where $I_0,I_1,I_2$ are the modified Bessel functions of order $0,1,2$, respectively.
In the homogeneous phase, $E>E_t=3/4$, the r.h.s. is equal to $\beta/2$, because the magnetization is zero.
This result is the same as the one obtained in Ref.~\cite{Patelli}.
Using formulas~\eqref{eq:AsyIntResp} we find the response of the magnetization 
\begin{equation}
	\delta m=\varphi_1(0)\frac{\lambda_1}{1-\lambda_1},
	\label{eq:HMFvariationHamiltonian}
\end{equation}
where $\varphi_1(0)=1$.
The dependence of $\lambda_1$ on energy $E$ for $E>E_t$ is shown in Fig.~\ref{fig:EqLambda} (dotted green line). 
As the transition temperature $E_t$ is approached from above, $\lambda_1 \to 1$. 

The Liouville-Nemuann series converges when the eigenvalue $\lambda_1$ is less than one in modulus.
Formula~\eqref{eq:Equilibrium} gives an eigenvalue larger than one in the inhomogeneous phase $E<E_t$.
As a consequence, the equilibrium distribution without any constraints is unstable in linear theory and
the response of magnetization diverges in the whole inhomogeneous phase.
This result can be physically justified because without the mass constraint the system could evaporate 
or collapse in a point, making the response undefined. This result was previously obtained in 
Ref.~\cite{InhomogeneousStabilityCampa}. The dotted green line in Fig.~\ref{fig:EqLambda} shows the 
eigenvalue as a function of $E$ also for $E<E_t$ where it takes values larger than one. 

\subsection{Equilibrium with the mass constraint}
The mass constraint imposes a different kernel of the integral~\eqref{eq:KernellMassOnly}.
The eigenvalue has the same eigenfunctions (sine and cosine), but it is described by
\begin{equation}
	\lambda_i=\beta\Phi[\varphi_if_{eq}']-\phi_0(q)\Phi[f_{eq}'].
\end{equation}
where $\phi_0(q)=m\cos(q)$.
The two eigenvalues $\lambda_{1,2}$ are
\begin{equation}
	\lambda_1=\beta\frac{I_1(\beta m)+\beta mI_2(\beta m)}{\beta mI_0(\beta m)}-\beta m^2,\quad \lambda_2=0.
	\label{eq:LMass}
\end{equation}
The dotted blue line in Fig.~\ref{fig:EqLambda} shows the eigenvalue in both the homogeneous and inhomogeneous phase.
Inserting it in formula~\eqref{eq:HMFvariationHamiltonian} we obtain a formula for the response of magnetization
which equals the ``canonical response"~\eqref{eq:ThermoCanonicalResp}.
In this case the eigenvalue $\lambda_1$ is equal to one only at the transition energy $E_t$.
Below and above this energy it is smaller than one, which ensures the existence of a finite asymptotic response 
in the canonical ensemble.

\subsection{Equilibrium with both energy and mass constraints}
The equilibrium case with both the energy and the mass constraint is described by the following eigenvalue equation 
\begin{equation}
	\lambda_1 =\beta\Phi[f'\varphi]+\delta\beta[\varphi]\Phi[f'\mathcal{H}_0]+\delta M[\varphi]\Phi[f']~.
	\label{eq:EigenEquationMassEnergy}
\end{equation}
Performing some algebraic manipulations, we obtain the formula for the eigenvalues
\begin{equation}
	\lambda_1=\beta\left(c-m^2\frac{1+2\beta^2c(c-m^2)}{1+2\beta^2m^2(c-m^2)}\right),\quad \lambda_2=0.
	\label{eq:LEnergyMassConstr}
\end{equation}
where
\begin{equation}
	c=\frac{I_1(\beta m)+\beta mI_2(\beta m)}{\beta mI_0(\beta m)},
\end{equation}
The full red line of Fig.~\ref{fig:EqLambda} shows the value of $\lambda_1$ for different energies.
That eigenvalue is again equal to one only at the transition energy $E_t$.
Moreover, the response of the magnetization~\eqref{eq:HMFvariationHamiltonian} gives the same 
expression as the ``microcanonical response"~\eqref{eq:MicroCanThermo}.

The full red line of Fig.\ref{fig:Resp} shows the response of magnetization in both the homogeneous 
and inhomogeneous phase. The response of the system with more constraints is smaller, consistently with the 
known result for the equilibrium response~\cite{InhomogeneousStabilityCampa,review3}.

The purple line in Figs.~\ref{fig:EqLambda} and~\ref{fig:Resp} reproduces the exact eigenvalue and the exact response
of magnetization derived Ref.~\cite{Yamaguchi:2012} using action-angle variables. As expected, the response 
which takes into account all the Casimir constraints gives an even smaller response.

\subsection{The Fermi-Dirac distribution}
\label{EigenQSS}
\begin{figure}[h!]
	\centering
	\includegraphics[width=75mm]{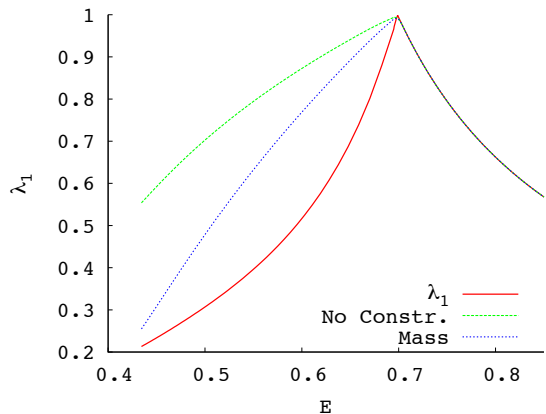}
	\caption{\small{
		Eigenvalue $\lambda_1$ for the Fermi-Dirac distribution~\eqref{eq:FermiDirac} without constraints
		(dotted green line), with mass constraint (dotted blue line) and with mass and energy constraint
		(full red line). The phase transition energy is $E_t \simeq 0.7$.
	}}
	\label{fig:FdLambda}
\end{figure}

\begin{figure}[h!]
	\centering
	\includegraphics[width=75mm]{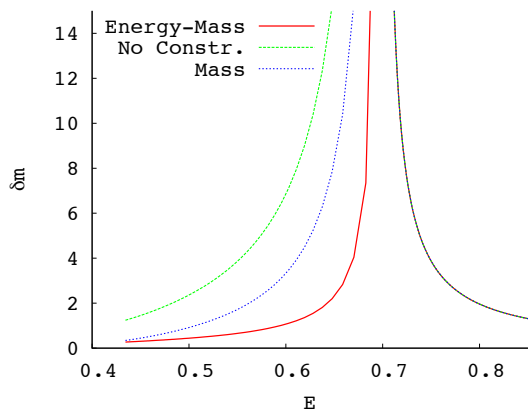}
	\caption{\small{
		The Response $\delta m$	for the Fermi-Dirac distribution without constraints
		(dotted green line), with mass constraint (dotted blue line) and with mass and energy constraint
		(full red line).
	}}
	\label{fig:FdResp}
\end{figure}

The Fermi-Dirac (FD) distribution is defined as
\begin{equation}
	f_{fd}(\mathsf{x})=\frac{1}{Z}\frac{1}{1+e^{\mathsf{x}}},
	\label{eq:FermiDirac}
\end{equation}
where $Z$ is the normalization.
This distribution is useful because it interpolates between different QSSs often studied in 
the framework of long-range interacting systems: the BG equilibrium and the Water-Bag states~\cite{review3}.
The eigenvalue equation uses the first derivative of the distribution with respect to its argument, which is
\begin{equation}
	f_{fd}'=-\frac{1}{4Z}\frac{1}{\cosh^2(\frac{\mathsf{x}}{2})}~.
\end{equation}
The eigenvalue equation for the FD distribution can be solved only numerically. Fig.~\ref{fig:FdLambda} shows the 
dependence of the eigenvalue $\lambda_1$ on energy $E$. It should reach the value $1$ at the transition energy $E_t\simeq0.7$,
but numerically (we use an iterative algorithm) we find a value slightly smaller than $1$.
It is intriguing that the zero constraint case gives an eigenvalue smaller than one also in the inhomogeneous phase,
at variance with the BG equilibrium case. This could be due to the fact that the FD state is more compact if compared 
with BG equilibrium and, therefore, it is probably more stable under evaporation.
Fig.~\ref{fig:FdResp} shows the response of magnetization $\delta m$ versus the energy $E$.
%% FIGURE

%% FIGURE

%-------------------------------------------------------------------------
\section{Dynamical evolution of the response}
\label{sec4:Numerics}

In the previous Section, we have determined analytically the response of the magnetization in the
$t \to \infty$ limit for the HMF model. Here, we want to show how the response evolves in time when the perturbing external
field is switched on. The analysis is purely numerical. We study both the BG equilibrium state and
the FD distribution.

In our simulations we use the weighted $N$-particle algorithm described in Ref.~\cite{Barre:2012}.
We prepare $N$ particles on a regular lattice of points in the $(q,p)$ phase-space and we associate 
a weight to each site corresponding to the chosen distribution, BG or FD.
The time evolution is realized using a symplectic integrator and every observable is evaluated using 
the weighted average. We have checked in the numerics that Vlasov dynamics conserves the support of every 
distribution at constant weight~\cite{Morrison:1987}.

The phase-space is compact in the $q$ direction, then errors depend on the lattice spacing.
In the $p$ direction we have another source of errors: for every distribution with a non compact 
support we have to use a cutoff on the weightless part of the momentum space.
We choose $p_{max}$ as the maximum value of the velocity spanned by the lattice and we adjust its
value in order to reduce the error. The optimal value of $p_{max}$ depends on the particular 
distribution considered.

The initial state is prepared at a given inverse energy scale $\sigma_0$. Solving an implicit equation 
we get the magnetization and the energy values associated with the chosen energy scale.
The perturbation is switched on at time $t_0>0$, after the system has relaxed to the stationary state.
We perturb with a field aligned along the direction of the spontaneous magnetization, in our case the $x$
direction.

There is not a well defined range of validity of linear response theory: whenever it gives a finite result
for a given observable, there will be always a sufficiently small value of the external field $h$ for 
which linear theory is a good approximation. In general we know empirically that in the homogeneous phase 
the amplitude of the perturbation must verify the relation $\sigma_0 h \ll 1$. In the inhomogeneous phase 
we have another scale, which is related to the non vanishing magnetization. It is quite reasonable to assume that
$h/m \ll 1$, in order to avoid that the external field gives a dominant contribution to the energy as 
compared with the magnetization. A first numerical verification of the validity of the linear regime is to
check that the variation of the magnetization is smaller than the magnetization itself $(h\delta m)/m \ll 1$
at all times for which the Vlasov equation is a good approximation of the finite $N$ dynamics.

\subsection{Equilibrium distribution}
\begin{figure}[h!]
	\centering
	\includegraphics[width=65mm]{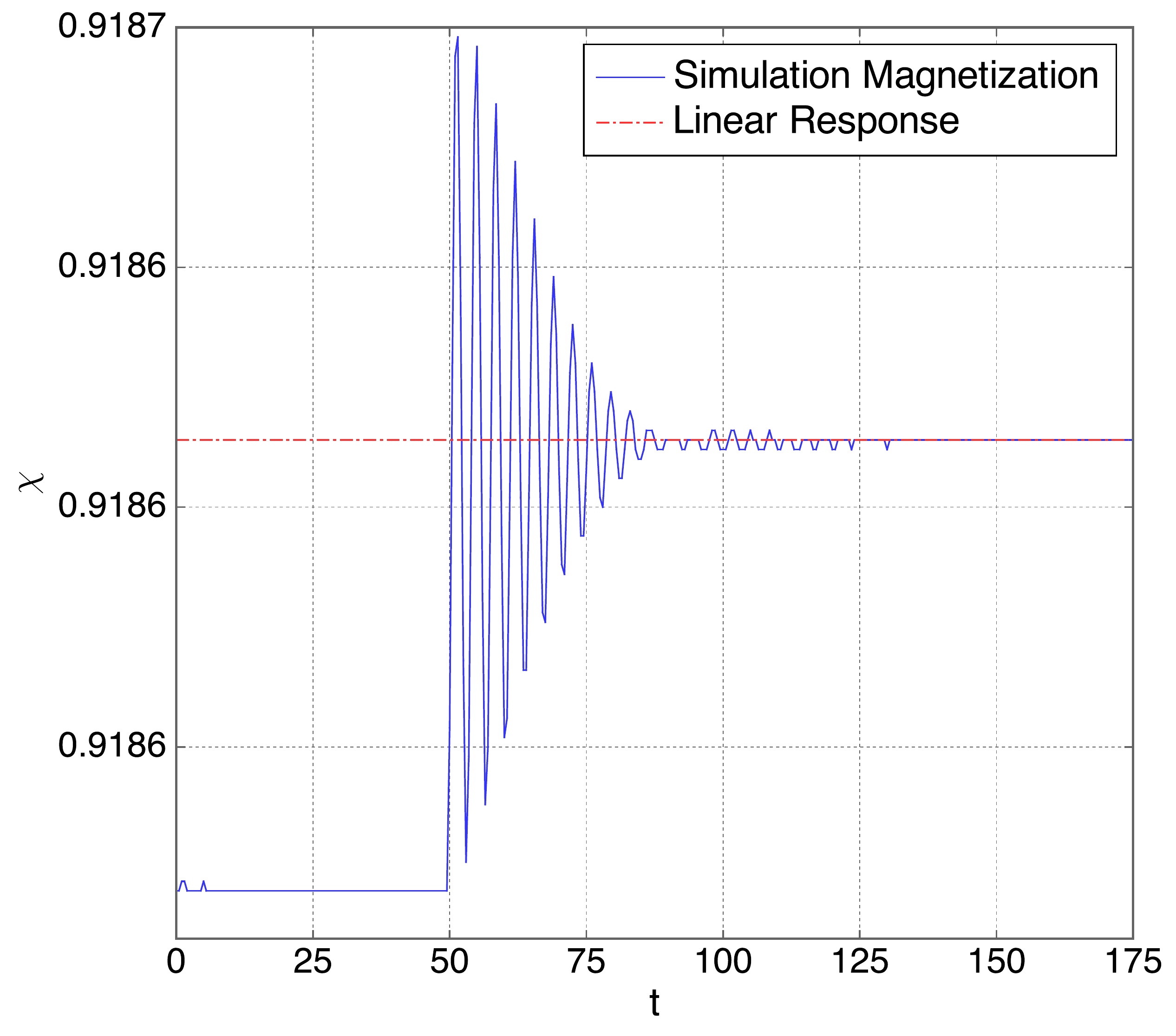}
	\caption{\small{Response of the magnetization $\delta m$ vs. time $t$ (full blue line) obtained in numerical 
	simulations of the HMF model perturbed
	around an inhomogeneous BG equilibrium state. Energy is $E=0.1495$ and magnetization $m=0.9185$, which corresponds 
	to the inverse energy scale $\sigma_0=\beta=7$. The external field is switched on at time $t_0=50$. The red dash-dotted 
	line is the theoretical prediction of linear response theory for asymptotic time.
	}}
	\label{fig:EqSimulations1}
\end{figure}

\begin{figure}[h!]
	\centering
	\includegraphics[width=65mm]{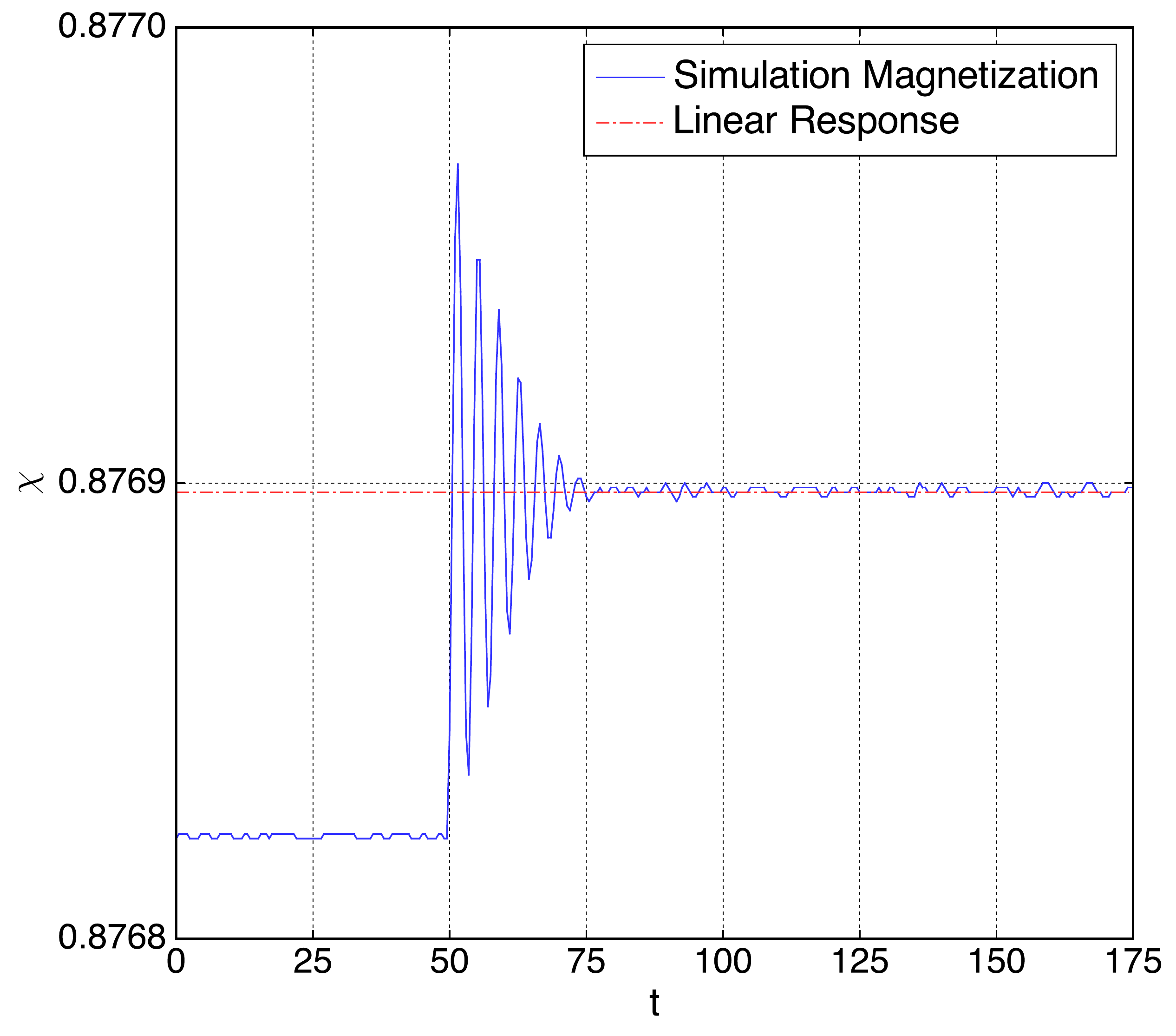}
	\caption{\small{Same as Fig.~\ref{fig:EqSimulations1} but with an energy $E=0.21559$ and magnetization $m=0.87682$, 
		which corresponds to an inverse energy scale $\beta=5$.
	}}
	\label{fig:EqSimulations2}
\end{figure}

We begin at time $t=0$ with an N-particle realization of the equilibrium BG distribution and we let it relax to
a stationary state before we apply the external field at time $t_0=50$. Figs.~\ref{fig:EqSimulations1} and~\ref{fig:EqSimulations2} 
display the results of simulations performed at different energies. In Fig.~\ref{fig:EqSimulations1} we show
the time evolution of the response of the magnetization (in blue) vs. time. At $t_0=50$, when the perturbation 
is applied, $\delta m$ begins to oscillate with a damping, a behavior that can be described within the theory 
of Landau damping~\cite{Landau:1946}. After some time, which depends mainly on the initial energy, the system 
relaxes to a stationary state whose magnetization is in good agreement with the theoretical predictions
of linear response theory given in Eq.~\eqref{eq:MagnEq} (red horizontal line).

Fig.~\ref{fig:EqSimulations2} shows a simulation at a higher energy. As in the previous case, after damping, the 
system relaxes to a state with a magnetization close to the value predicted by linear response theory.

%% FIGURE

\subsection{Fermi-Dirac distribution}
\begin{figure}[h!]
	\centering
	\includegraphics[width=65mm]{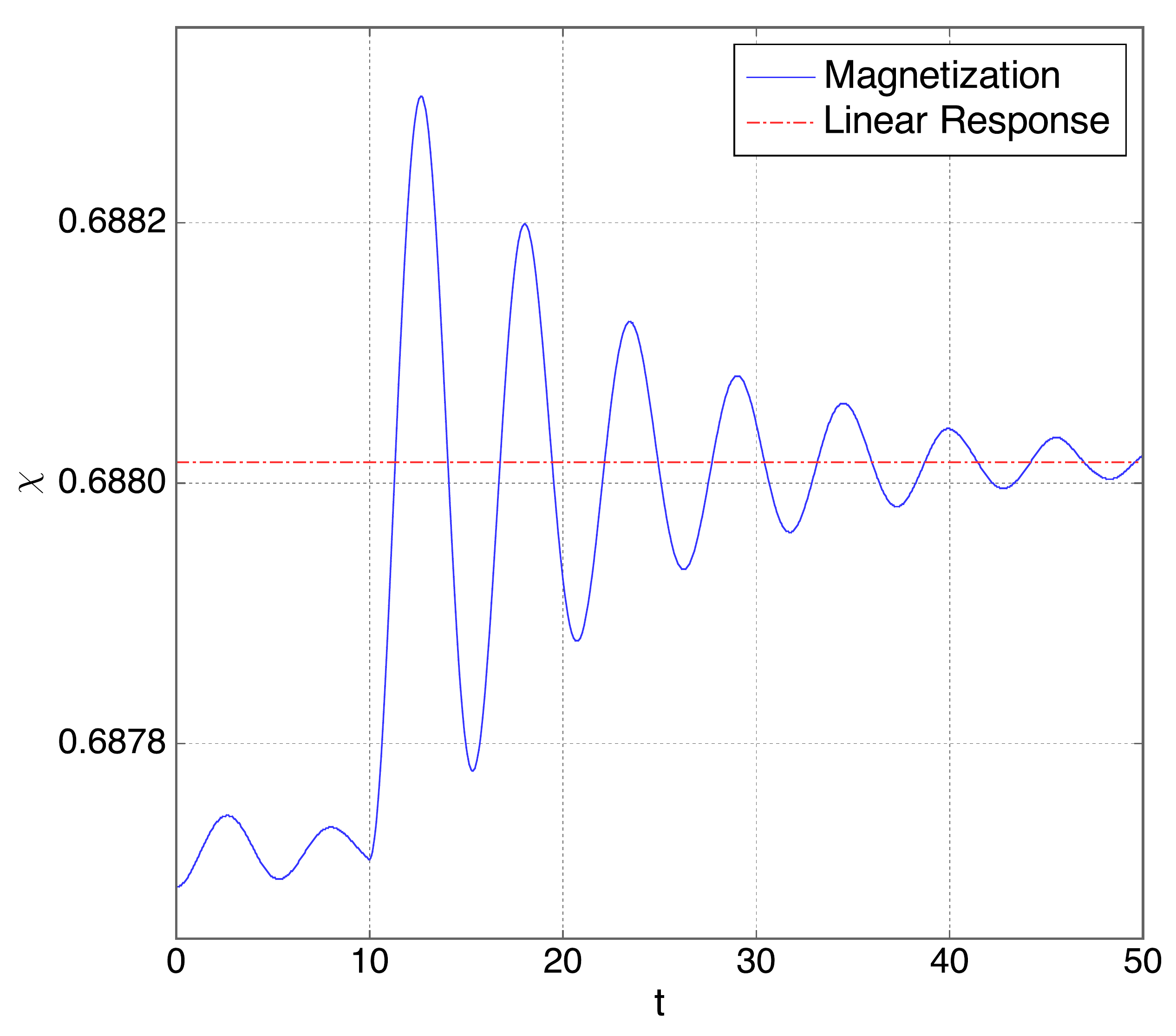}
	\caption{\small{Response of the magnetization $\delta m$ vs. time $t$ (full blue line) obtained in numerical 
	simulations of the HMF model perturbed around an inhomogeneous Fermi-Dirac state. 
	Energy is $E=0.4481$ and magnetization $m=0.6877$, which corresponds 
	to the inverse energy scale $\sigma_0=\beta=8$. The external field is switched on at time $t_0=10$. The red dash-dotted 
	line is the theoretical prediction of linear response theory for asymptotic time.
	}}
	\label{fig:FdSimulations1}
\end{figure}

\begin{figure}[h!]
	\centering
	\includegraphics[width=65mm]{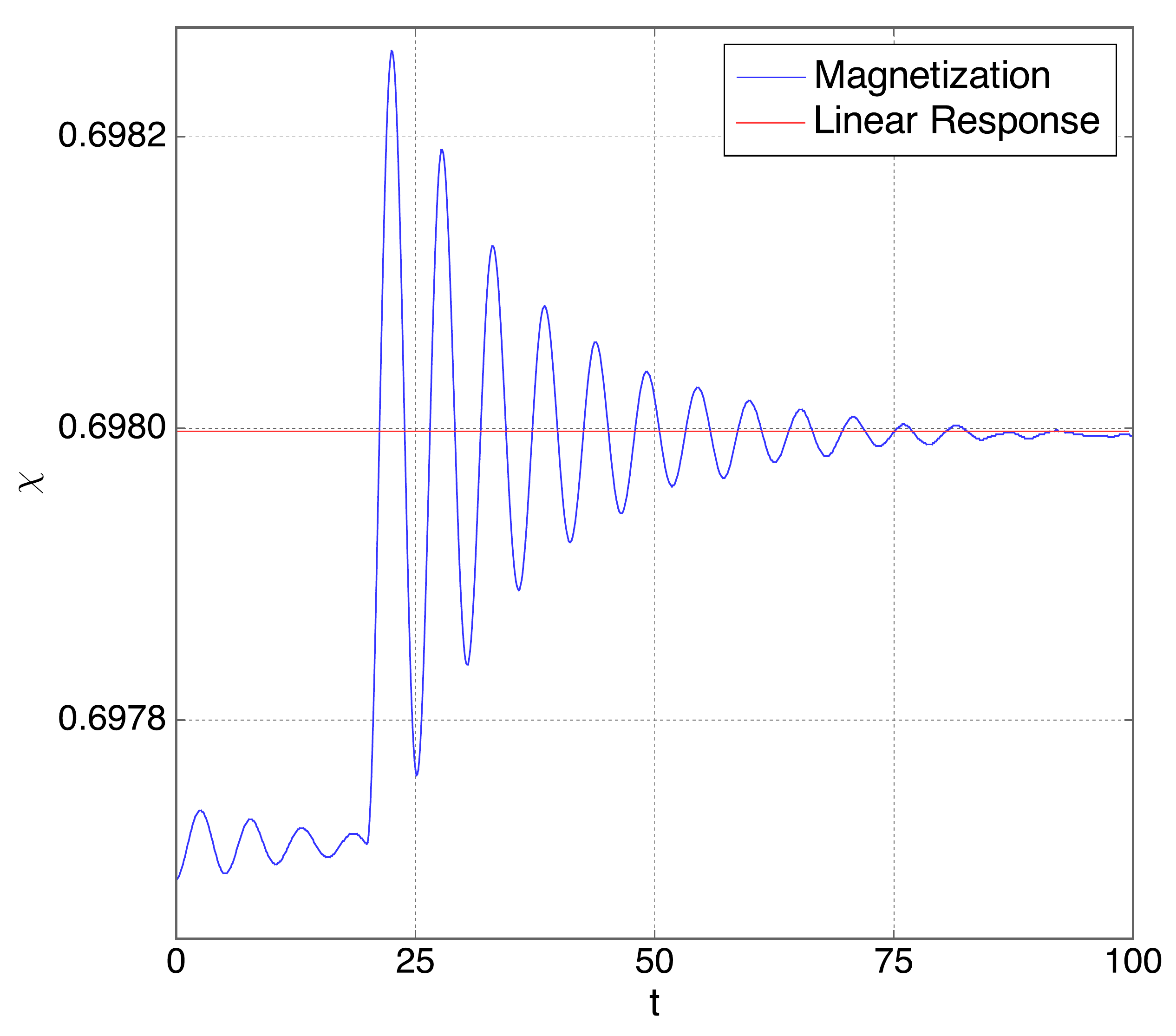}
	\caption{\small{Same as Fig.~\ref{fig:FdSimulations1}, but with energy $E=0.4401$ and a magnetization $m=0.6977$,
		which corresponds to the inverse energy scale $\sigma_0=\beta=9$. The external field is here switched on at
		time $t_0=20$.
	}}
	\label{fig:FdSimulations2}
\end{figure}

Figs.~\ref{fig:FdSimulations1} and~\ref{fig:FdSimulations2} show the comparison between the numerical simulations
(blue lines) and the theoretical predictions for the asymptotic magnetization (red horizontal line) for the Fermi-Dirac 
distribution~\eqref{eq:FermiDirac}.

%% FIGURE

%% FIGURE

%%%%%%%%%%%%%%%%%%%%%%%%%%%%%%%%%%%%%%%%%%%%%%%%%%%%%%%%%%%%%%%%%%%%%%%%%%%%%%%%%%%%%%%%
\section{Conclusions}
\label{Conclusions}
%%%%%%%%%%%%%%%%%%%%%%%%%%%%%%%%%%%%%%%%%%%%%%%%%%%%%%%%%%%%%%%%%%%%%%%%%%%%%%%%%%%%%%%%
We have studied linear response to a small external perturbation for long-range interacting $N$-particle systems 
trapped in quasistationary states (QSS)~\cite{Patelli}. QSS are described by stable and stationary solutions of the Vlasov equation, 
in the $N \to \infty$ limit when ``collisions" are rare and negligible. We have studied this problem in the context
of the Hamiltonian Mean Field (HMF) model~\cite{Ruffo:1995}, a paradigmatic example of long-range interacting system. The model
is characterized by the presence of both homogeneous and inhomogeneous stationary states.
While homogeneous QSS can be described following the work of Landau~\cite{Landau:1946}, inhomogeneous QSS present more difficulties 
for their analytical treatment. They show a coupling among all the ``modes" of the system, as discussed above.
In the case where the mean-field effective potential generates integrable motion, an exact linear
response formula can be derived using a transformation to action-angle variables~\cite{Yamaguchi:2012}. This is the case for
the HMF model. However, many long-range interacting systems of physical interest, like self-gravitating systems and
plasmas, do not fall into the category of integrable systems.
In this paper, we have derived an approximate linear response formula for generic, i.e. non integrable, systems endowed
with a finite, but arbitrary, number of integrals of motion and Casimir invariants.
The linear response formula that we have obtained describes also the infinite time limit of the system. Therefore, 
it gives a description of the new quasistationary state reached asymptotically after the application of the perturbation,
whenever the initial unperturbed state is stationary stable.
We have applied this formalism to the HMF model, which is an integrable system, in order to compare our
approximate result with the exact one obtained in Ref.~\cite{Yamaguchi:2012}.
We have found a good agreement of the linear response of magnetization when we impose the constraints of mass
and energy conservation away from transitions points from inhomogeneous to homogeneous states.
We have also compared the predictions of the theory with simulations performed at finite $N$.
The method we have devised can be used also for non integrable models and we look forward to this application.

\section{Acknowledgments}
\label{Ack}

We acknowledge J. Barr\'e, P. H. Chavanis, C. Nardini, S. Ogawa and Y. Yamaguchi for useful discussions. We thank the Galileo Galilei Institute for Theoretical Physics (Florence) for the hospitality and INFN for partial support during the completion of this work

\appendix{
%%%%%%%%%%%%%%%%%%%%%%%%%%%%%%%%%%%%%%%%%%%%%%%%%%%%%%%%%%%%%%%%%%%%%%%%%%%%%%%%%%%%%%%%
\section{Considerations on Energy-Casimir}
\label{app4:Energy-Casimir}
%%%%%%%%%%%%%%%%%%%%%%%%%%%%%%%%%%%%%%%%%%%%%%%%%%%%%%%%%%%%%%%%%%%%%%%%%%%%%%%%%%%%%%%%
The theory of Casimirs allows one to obtain stationary solutions of the Vlasov equation from a variational 
principle~\cite{Holm:1984,Morrison:1987,StabilityShun,InhomogeneousStabilityCampa}. It is also useful to compare the functional form of different distributions.

Calling $c(y)$ a generic invertible and differentiable function, the Casimir functional is 
\begin{equation}
	\mathcal{C}[f]=\int c(f(x,v))dxdv,
\end{equation}
The function $c$ is referred to as the generator of the Casimir. The Vlasov equation describes isolated systems, 
therefore, it conserves both energy and mass. Therefore, it is possible to write the variational equation
\begin{equation}
	\max_{f}\Big\{\mathcal{C}[f] \Big| \int f(x,v)\mathrm{E}(x,v)=E,\int f(x,v)=1\Big\},
\end{equation}
and its extremal distribution
\begin{equation}
	f(q,p)=(c')^{-1}(\sigma \mathcal{H}(q,p)),
	\label{eq:DistrFunc}
\end{equation}
where $\sigma$ is an inverse energy scale.
The function $c'$ is the first derivative of the function and the label $-1$ stands for the inverse.
Distribution~\eqref{eq:DistrFunc} is stationary as stated by Jeans, because it depends only on functions such as the Hamiltonian.
We denote stationary distributions with the $0$ subscript.
The value of the Casimir can be written as a function of the distribution
\begin{equation}
	\mathcal{C}[f_0]=\sigma_0\langle \mathcal{H}_0\rangle_0+\langle \int^{\sigma_0 \mathcal{H}_0}_{z^*}
	\frac{f_0(y)}{f_0(\sigma_0 \mathcal{H}_0)}dy\rangle_0
\end{equation}
where $f_0(z^*)=0$.
This variational approach is related to a kind of stability analysis because the solution must maximize 
the Casimir functional~\cite{Morrison:1987}.

Inverting this procedure, we consider a known {\em initial} distribution function $f_0$ which can be generated by a 
Casimir with generator $c(y)$. Moreover, let us consider that the {\em final} state satisfies the Casimir variational 
principle, but with a different generator $t(y)$. Linear response theory requires that in the limit of vanishing 
perturbation $h \to 0$ the Casimirs coincide, hence we develop the Casimir $\mathcal{T}$ around the unperturbed 
state $f_0$
\begin{equation}
	\mathcal{T}[f]\to \int \,c(f_0)+h\int c'(f_0)\delta f+h\int g(f_0)+\mathcal{O}(h^2).
\end{equation}
The function $g$ represents the variation of the generating function of the Casimir due to the perturbation.
Using equation~\eqref{eq:DistrFunc}, the first part of that variation is given by the variation of the 
Hamiltonian $\int \mathcal{H}_0\delta f$ and it is zero when the system conserves the energy.
Therefore, the variation of the Casimir depends only on the function $g$, which corresponds to a variation of 
the generator of the Casimirs induced by the perturbation. Indeed, the presence of a variation of the generator 
implies a different functional form of the final state compared to the initial one.
The corresponding equation arising from the variational principle reads
\begin{equation}
	\delta f c''(f_0) - \delta \sigma \mathcal{H}_0 - \sigma_0 \delta  \mathcal{H} - \delta M + g(f_0) = 0.
	\label{eq:SecondVariationaPrinciple}
\end{equation}
This equation relates the variation of the distribution function $\delta f$ to the variation of the Hamiltonian 
and corresponds to the Duhamel formula at infinite times.

From a numerical point of view a variation of the generator of the Casimir generator induces a variation in the 
value of the Casimir in time, as can be easily checked.

Moreover, the variation of the energy in perturbed systems, when the perturbation is instantaneously switched on, is zero. 
In analogy, the variation of the generator of the Casimir is zero with perturbations instantaneously switched on.

%%%%%%%%%%%%%%%%%%%%%%%%%%%%%%%%%%%%%%%%%%%%%%%%%%%%%%%%%%%%%%%%%%%%%%%%%%%%%%%%%%%%%%%%%
\section{Asymptotic time limit of $J_\omega$}
\label{app4:TheSecondIntegral}
%%%%%%%%%%%%%%%%%%%%%%%%%%%%%%%%%%%%%%%%%%%%%%%%%%%%%%%%%%%%%%%%%%%%%%%%%%%%%%%%%%%%%%%%%
We have to evaluate the integral~\eqref{eq:ManifoldIntegral} in the limit $\omega \to 0$.
This limit is not trivially zero, but its value is determined by terms arising from singularities
in manifolds where the Liouville operator is zero.

The motion of a particle in an external potential $\psi(x)$ can be bounded or unbounded.
One of the main differences between these two kinds of motion comes from the spectral properties of the generator of the dynamics.
For instance, the manifold in which the motion changes from bounded to unbounded could show a "zero" dynamics, i.e.
the Liouville operator constrained on such manifold is strictly zero.
Therefore, the limit of equation~\eqref{eq:ManifoldIntegral} gains a divergence on such manifolds
and could introduces a new term in the eigenvalue equation.
For example, the dynamics of the unperturbed HMF model is the dynamics of the pendulum which shows stable and unstable points, 
generating a separatrix between bounded and unbounded motion.
Along the separatrix the frequency of the bounded motion diverges and the Liouville operator is zero.

We have to evaluate the following integral
\begin{equation}
	\mathcal{J}_\omega=\int dxdvf_0'(\mathcal{H}_0)\varphi(x)\frac{\omega}{\omega -\imath\mathcal{L}_0}e^{\imath kx},
	\label{eq:InitiaIntegral}
\end{equation}
in the $\omega\to0$ limit, where the operator
\begin{equation}
	\mathcal{L}_0g(x,v)=-v\partial_x g +m\partial_x\psi(x)\partial_vg.
\end{equation}
is the Liouville operator of the unperturbed dynamics.
The unperturbed mean-field potential $m$ is the modulus of the magnetization of the HMF model.
First of all, we separate the two regions of positive and negative velocities. The Liouville operator 
in the two regions has the symmetry property
\begin{equation}
	\mathcal{L}_0^+=-\mathcal{L}_0^-.
\end{equation}

Let us now introduce the following transformation of coordinates
\begin{equation}
	(x,v)\rightarrow(x,E),\qquad mE=\frac{v^2}{2}-m\psi(x).
\end{equation}
The Liouville operator becomes
\begin{equation}
	\mathcal{L}_0^+=\sqrt{m}\sqrt{2(E+\psi)}\partial_x\psi\partial_E.
\end{equation}
The integral~\eqref{eq:InitiaIntegral} becomes
\begin{equation}
	\mathcal{J}_\omega=\sqrt{m}\int dxdE \frac{f_0'(E+2\psi)}{\sqrt{2(E+\psi)}}\varphi(x)\frac{\omega^2}{\omega^2 
	+ \mathcal{L}_0^2}e^{\imath kx}.
\end{equation}
The square of the operator acting on a function $g(x)$ of the spatial variable $x$ gives
\begin{equation}
	\mathcal{L}_0 g(x)=0,\qquad \mathcal{L}_0^2g(x)=m(\partial_x\psi(x))^2g(x),
\end{equation}
and, consequently, the integral looses its operatorial form
\begin{equation}
	\mathcal{J}_\omega=\sqrt{m}\int dx\frac{\varphi (x)e^{\imath kx}}{1 + m\left(\partial_x\psi/\omega\right)^2}\int dE \frac{f_0'(m(E+2\psi))}{\sqrt{2(E+\psi)}}.
\end{equation}
This integral does not depend on $\omega$ only in the manifold where the derivative of the 
mean-field potential $\partial_x\psi$ is zero.
When the Lebesgue measure of that manifold is not zero, integral~\eqref{eq:InitiaIntegral} gives a contribution 
to the eigenvalue equation.

For instance, in the homogeneous case, the mean-field potential is zero and the integral above converges to zero
in the limit of vanishing frequencies, because the manifold of vanishing mean-field force is of zero measure.
This is consistent with the results shown in Section~\ref{sec4:Asymptotics}.

}

%
% BibTeX users please use
% \bibliographystyle{}
% \bibliography{}

\begin{thebibliography}{}

\bibitem{review3} 
A. Campa, 
T Dauxois, 
S. Ruffo, 
Phys. Rep., \textbf{480}, 
57 (2009).

\bibitem{Leshouches}
T. Dauxois, S. Ruffo and L. Cugliandolo (Eds.), {\it Long-range interacting systems}, 
Oxford University Press (2010).

\bibitem{Yamaguchi:2004}
Y. Y. Yamaguchi,
J. Barr\'e,
F. Bouchet,
T. Dauxois,
S. Ruffo,
Physica A, \textbf{337},
36 (2004).

\bibitem{Balescu:1997} 
R. Balescu, 
\textit{Statistical Dynamics: Matter Out of Equilibrium}, 
(Imperial College Press, London, 1997).

\bibitem{Nicholson:1992} 
D. R. Nicholson, 
\textit{Introduction to Plasma Physics}, 
(Krieger Publishing Company, Florida, 1992).

\bibitem{Braun:1977}
W. Braun,
K. Hepp, 
Comm. Math. Phys., \textbf{56}, 
101 
(1977).

\bibitem{Patelli} 
A. Patelli, 
C. Nardini, 
S. Gupta, 
S. Ruffo, 
Phys. Rev. E, \textbf{85},
021133
(2012).

\bibitem{Yamaguchi:2012} 
S. Ogawa, 
Y. Y. Yamaguchi, 
Phys. Rev. E, \textbf{85} 
061115
(2012).

\bibitem{Chavanis:2013}
P. Chavanis,
EPJ Plus, \textbf{128}, 
38
(2013).

\bibitem{Landau:1946} 
L. Landau,
J. Phys. USSR, \textbf{10}, 
25
(1946).

\bibitem{Nardini1}
C. Nardini,
S. Gupta,
S. Ruffo,
T. Dauxois,
F. Bouchet,
J. Stat. Mech, \textbf{L01002}
(2012).

\bibitem{Nardini2}
C. Nardini,
S. Gupta,
S. Ruffo,
T. Dauxois,
F. Bouchet,
J. Stat. Mech, \textbf{P12010}
(2012).

\bibitem{Barre:2012}
J. Barr\'{e}, 
A. Olivetti,
Y. Y. Yamaguchi,
J. Phys. A, \textbf{45},
1 (2012).

\bibitem{Morrison:1987} 
P. J. Morrison,
Z. Natur., {\bf 42a}, 
1115 (1987).


\bibitem{Spohn:1982}
J. Messer,
H. Spohn,
J. Stat. Phys., \textbf{29},
561 (1982).


\bibitem{Inagaki:1993}
S. Inagaki,
T. Konishi,
Pubbl. Astro. Soc. J., \textbf{43},
733 (1993).


\bibitem{Ruffo:1995} 
M. Antoni,
S. Ruffo, 
Phys. Rev. E, {\bf 52},
2361
(1995).

\bibitem{Jeans:1919} 
J. H. Jeans, 
\textit{Problems of Cosmogony and Stellar Dynamics}, 
(Cambridge University Press, Cambridge, 1919).

\bibitem{KuboJPSJ:1957} 
R. Kubo, 
J. Phys. Soc. Japan, \textbf{12}, 
570
(1957).


\bibitem{Pakter:2013} 
R. Pakter,
Y. Levin, 
J. Stat. Phys., {\bf 150},
531 (2013).

\bibitem{ReedSimon}
M. Reed,
B. Simon, 
\textit{Methods of modern mathematical physics. II. Fourier analysis, self-adjointness},
(Academic Press, 1975).

\bibitem{KatoPertSeries}
T. Kato,
\textit{Perturbation Theory for Linear Operators},
(Springer-Verlag,1980).


\bibitem{Whittaker}
E.T.Whittaker,
G.N.Watson, 
\textit{A course of modern analysis},
(Cambrige University Press, 1915).


\bibitem{Penrose:1959} 
O. Penrose, 
Phys. of Fluid,\textbf{3}, 
2
(1959).

\bibitem{StabilityShun}
S. Ogawa,
Phys. Rev. E, \textbf{87},
062107
(2013).

\bibitem{Holm:1984} 
D. D. Holm, 
J. E. Marsden, 
T. Ratiu, 
A. Weinstein, 
Phys. Rep., {\bf 123}, 1
(1985).


\bibitem{Kac:1959}
M. Kac,
Phys. of Fluids, \textbf{2}, 8
(1959).


\bibitem{InhomogeneousStabilityCampa}
A. Campa,
P. Chavanis,
J. Stat. Mech., \textbf{P06001}
(2010).




\end{thebibliography}

\end{document}